\documentclass[pra,twocolumn]{revtex4}
\usepackage{amssymb}
\usepackage{amsmath}
\usepackage{graphicx}
\usepackage{dcolumn}
\usepackage[center]{subfigure}
\usepackage{booktabs}
\usepackage{epstopdf}
\usepackage{xcolor}

\begin{document}

\title{Pancake-shaped vortex droplets in dipolar molecular BECs}
\author{Zibin Zhao$^{1,2}$}
\thanks{These authors contributed equally to this work.}
\author{Xinyi Tang$^{1}$}
\thanks{These authors contributed equally to this work.}
\author{Tianmiao Zhang$^{1}$}
\author{Zhaopin Chen$^{3}$}
\author{Huanbo Luo$^{1}$}
\email{huanboluo@fosu.edu.cn}
\author{Guihua Chen$^{4}$}
\author{Bin Liu$^{1}$}
\author{Boris A. Malomed$^{1,5,6}$}
\author{Yongyao Li$^{1}$ }
\email{yongyaoli@gmail.com}
\affiliation{$^1$School of Physics and Optoelectronic Engineering, Foshan University,
Foshan 528000, China\\
$^2$Moscow Institute of Physics and Technology, Dolgoprudny 141701, Moscow Region, Russia\\
$^3$Physics Department and Solid-State Institute, Technion, Haifa 32000, Israel\\
$^{4}$Department of Electronic Engineering, Dongguan  University of Technology, Dongguan, 52300, China\\
$^5$Department of Physical Electronics, School of Electrical Engineering,
Faculty of Engineering, Tel Aviv University, Tel Aviv 69978, Israel\\
$^6$Instituto de Alta Investigaci\'{o}n, Universidad de Tarapac\'{a},
Casilla 7D, Arica, Chile }

\begin{abstract}
Anisotropic interactions profoundly affect topological excitations in
quantum fluids. Motivated by recent advances in the studies of
microwave-shielded polar molecules, we introduce self-trapped modes in the
form of pancake-shaped quantum droplets (QDs) with embedded vorticity, which
are maintained by strongly anisotropic dipole-dipole interactions. The
stability region of singly charged ($S=1$) vortices nearly coincides with
that of the ground-state QDs ($S=0$), demonstrating the robustness of the
vortex states. The strong anisotropy of the system splits the core (pivot)
of vortex QDs with $S=2$ into separated unitary ones. The angular momentum
and stability of the state with $S=2$ are affected by the separation between
the unitary cores. Higher-charge vortex QDs with $S>2$ are stable too, for
sufficiently large particle numbers and inter-core separations. On the other
hand, bound vortex-antivortex pairs with $S=\pm 1$ are unstable. Head-on
collisions between the vortex QDs exhibit distinct regimes, including
rebound, merger, and fragmentation. Tuning the cylindrically symmetric component of the dipolar interaction reveals a pronounced sign-dependent response: positive tuning preserves the self-bound vortex, whereas negative tuning drives expansion and fragmentation. The results demonstrate that the
microwave-dressed molecular QDs offer a robust platform for the realization
of self-trapped vortex states, demonstrating how the strong anisotropy
reshapes their structure, stability, and dynamics.
\end{abstract}

\maketitle

The observation of self-bound quantum droplets (QDs) in ultracold polar
molecular condensates has established molecules as a new means for the
realization of dipolar quantum fluids \cite{S-Zhang}. External electric and
microwave fields, along with microwave shielding, enable precise control of
the strength, sign, and spatial anisotropy of the dipole-dipole interaction
(DDI), while suppressing collisional losses \cite%
{Anderegg,PRX-Quantum,W-Yuan,IRli,DJwang,Schindewolf,Bigagli,Karman,Jlin,
Schindewolf2026,Jozwiak2026,ZZeng}%
. In particular, microwave dressing can strongly break the rotational
symmetry \cite{PRX-Quantum,Baillie,FDeng,FDeng-tWO,TArnone,Chen-FLR}.
Molecular QDs remain stable for over $100\,\mathrm{ms}$ in the regime of
strong interactions, and the underlying molecular BECs can persist in the
course of seconds \cite{S-Zhang,DJwang,Bigagli-BEC}. This combination of the
tunability and longevity provides a favorable setting for the implementation
of quantum simulations and data processing \cite%
{DeMille2002,Micheli2006,Brennen2007,Cornish2024,Karra2016,Li2023,Bao2023},
few-body bound states \cite{Efimov-dipoles,XYChen,BWunsch,Quemener,Tkarman},
supersolids \cite%
{Pollet,He-supersolid,Schmidt-supersolid,Capogrosso,WZhang,KPSchmit}, and
topological excitations, including vortices, in self-bound molecular QDs
\cite{Deng-vortex,Deng-SOC-vortex,SvSyzranov,NRCooper}.

Vortex states are highly sensitive to anisotropy \cite%
{Mayteevarunyoo-aniso,Zhang-semidiscrete,Terhalle-aniso,Chen-aniso,Ge-nonlocal,Liao-Zeeman,Malomed-vortex-review}%
. In particular, in dipolar atomic condensates, vortex QDs have been found
with the vortex axis directed perpendicular to the dipole polarization,
thereby strongly breaking the rotational symmetry of the system, whereas
vortices aligned with the dipole direction are unstable, hence the
anisotropy plays a decisive role in the vortex stabilization \cite%
{GLi-2D,GLi-3D,ACidrim-unstable}. In the studies of molecular QDs, where the
DDI can itself be strongly anisotropic, a natural question is how the
anisotropy affects properties of vortex modes, including their symmetry and
structure. An extended question is whether vortex QDs with higher values of
the topological charge, which are usually vulnerable to instability against
azimuthal splitting, can be stabilized in the symmetry-breaking system.

In this work, we investigate pancake-shaped vortex QDs in the model of
dipolar molecular condensates, where the enhanced spatial anisotropy of the
dual-microwave-dressed DDI produces the oblate droplet morphology. The
stability region of the singly charged ($S=1$) vortex QD nearly coincides
with that of the ground-state (zero-vorticity) QDs, demonstrating the
robustness of the vortex modes in this system. For higher-order vorticities,
the broken rotational symmetry leads to splitting of vortex pivots, rather
than keeping a single one, with the stability strongly affected by the
separation between the pivots. We also identify the angular momentum of
vortices in the anisotropic system and examine collision dynamics of the
vortex QDs.

\begin{figure}[tbp]
{\includegraphics[width=0.95\columnwidth]{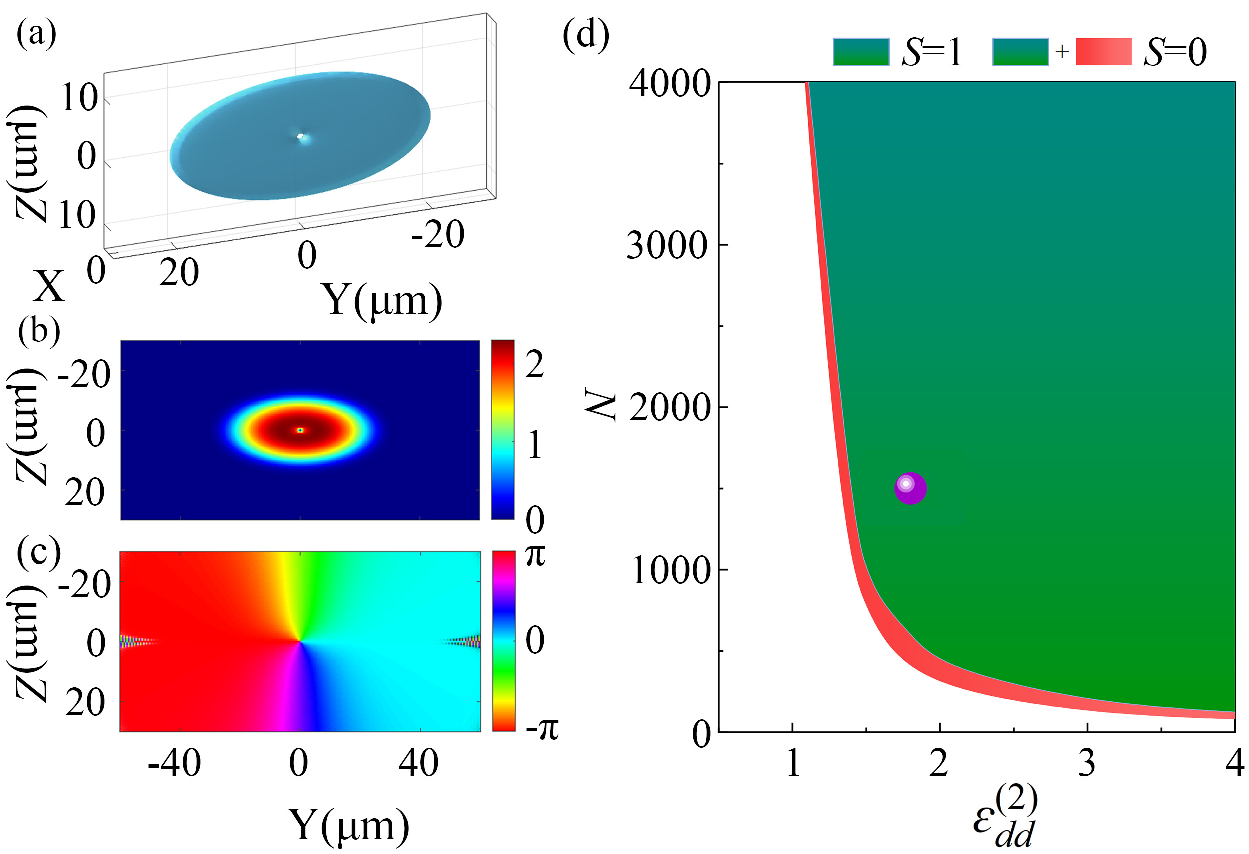}}
\caption{(a) A 3D isosurface plot, alongside the corresponding 2D (b)
density and (c) phase distributions in the $X=0$ plane, for the vortex QD
with $S=1$, $N=1500,$ and $\protect\epsilon _{\mathrm{dd}}^{(2)}=1.8$. (d) The
stability diagram for the QDs with $S=0$ (the ground state) and $S=1$. The
stability area for the $S=1$ QDs is shaded green, whereas the ground state,
with $S=0$, is stable in the combined green and orange domains. The purple
circle refers to the vortex solution shown in (a)--(c).}
\label{fig1}
\end{figure}

We consider a Bose-Einstein condensate of dipolar bialkali molecules dressed
by two microwave fields. The dressing configuration consists of a linearly
polarized field oscillating along the $Z$ axis, referred to as the $\pi $%
-field, and an elliptically polarized field rotating in the transverse $%
(X,Y) $ plane, referred to as the $\sigma $-field \cite{S-Zhang}. The
combined action of these fields gives rise to an anisotropic DDI between
molecules, with the interaction potential
\begin{equation}
U_{dd}(\mathbf{R})=\frac{3G_{s}}{4\pi R^{3}}\left[ \epsilon
_{\mathrm{dd}}^{(0)}(1-3\cos ^{2}\theta )+\sqrt{3}\epsilon _{\mathrm{dd}}^{(2)}\sin ^{2}\theta
\cos 2\phi \right] ,  \label{DDI}
\end{equation}%
where $R=|\mathbf{R}|$, with $\theta $ and $\phi $ being the polar and
azimuthal angles of $\mathbf{R}$, respectively \cite{PRX-Quantum}. The first
term in Eq. (\ref{DDI}) represents the cylindrically symmetric component of
the dressed interaction, whereas the second term explicitly breaks the
rotational symmetry in the transverse plane. The corresponding dimensionless
interaction strengths are defined as $\epsilon _{\mathrm{dd}}^{(0)}=a_{\mathrm{d0}}/a_{s}$ and
$\epsilon _{\mathrm{dd}}^{(2)}=a_{\mathrm{d2}}/a_{s}$, where $a_{s}$ is the effective $s$-wave
scattering length, while $a_{\mathrm{d0}}$ and $a_{\mathrm{d2}}$ characterize the strengths of
the two DDI components \cite{S-Zhang,PRX-Quantum,W-Yuan}. In particular, the
anisotropic component is controlled by ellipticity $\xi $ of the $\sigma $%
-field, according to $\epsilon _{\mathrm{dd}}^{(2)}\propto \sin (2\xi )$, providing a
direct control of the transverse anisotropy \cite{S-Zhang}. Coefficients $%
\epsilon _{\mathrm{dd}}^{(0)}$ and $\epsilon _{\mathrm{dd}}^{(2)}$ can be tuned across
positive, zero, and negative values by varying the dressing configuration.

The condensate dynamics are governed by the extended Gross-Pitaevskii
equation
\begin{gather}
i\hbar \frac{\partial \Psi }{\partial T}={}\left[ -\frac{\hbar ^{2}}{2m}%
\nabla ^{2}+\gamma _{\mathrm{LHY}}|\Psi |^{3}+G_{s}|\Psi |^{2}\right] \Psi
\notag \\
+\Psi \int U_{dd}(\mathbf{R}-\mathbf{R}^{\prime })\left\vert \Psi (\mathbf{R}%
^{\prime })\right\vert ^{2}d\mathbf{R}^{\prime },  \label{GP}
\end{gather}%
where $m$ is the molecular mass, $G_{s}=4\pi \hbar ^{2}a_{s}/m$ is the
contact-interaction coefficient, and $U_{\mathrm{dd}}$ is potential (\ref{DDI}). The
coefficient of the Lee-Huang-Yang term in Eq. (\ref{GP}) is
\begin{equation}
\gamma _{\mathrm{LHY}}=\frac{32G_{s}a_{s}^{3/2}}{3\sqrt{\pi }}\mathrm{Re}\!%
\left[ \mathcal{Q}_{5}\left( \epsilon _{\mathrm{dd}}^{(0)},\epsilon
_{\mathrm{dd}}^{(2)}\right) \right] ,  \label{gamma-QF}
\end{equation}%
with $\mathcal{Q}_{5}=\int [1+G_{s}\tilde{U}(\mathbf{k})]^{5/2} d\Omega/4\pi$.
Here, $G_s\tilde{U}(\mathbf{k})$ denotes the Fourier transform of $U_{\mathrm{dd}}(%
\mathbf{r})$ (See Ref. ~\cite{SRc} for its explicit form), and the function $%
\mathcal{Q}_{5}$ may be complex in part of the parameter space. However,
only its real part is retained in the numerical calculations, following the
widely adopted assumption \cite%
{PRA-1-img,PRA-2-img,ARPLima-low-lying,ARPLima-QF}. The total molecula number, $N=\int |\Psi |^{2}d^{3}\mathbf{R} $, is a dynamical invariant of
Eq. (\ref{GP}), if $\mathcal{Q}_{5}$ is taken as the real coefficient.

Stationary solutions are sought for as $\Psi (\mathbf{R},T)=\psi (\mathbf{R}%
)\exp (-i\mu T/\hbar )$, where $\mu $ is the chemical potential and $\psi (%
\mathbf{R})$ is the stationary wave function.

In this paper, we take the physical parameters as those of the $\text{Na}%
\text{Cs}$ molecules, with $a_{s}=2100a_{0}$ ($a_{0}$ is the Bohr radius),
using the norm (total number of particles) $N$ and the dipolar anisotropy $%
\epsilon _{\mathrm{dd}}^{(2)}$ as primary control parameters. To focus on effects of
anisotropy, we firstly fix $\epsilon _{\mathrm{dd}}^{(0)}=0$ and vary $\epsilon _{\mathrm{dd}}^{(2)}$
while keeping $\epsilon _{\mathrm{dd}}^{(2)}>0$. This parameter regime can be
realized experimentally by appropriately tuning the dual-microwave fields
\cite{S-Zhang,PRX-Quantum,W-Yuan,DJwang}. 

\begin{figure}[tbp]
{\includegraphics[width=1.01\columnwidth]{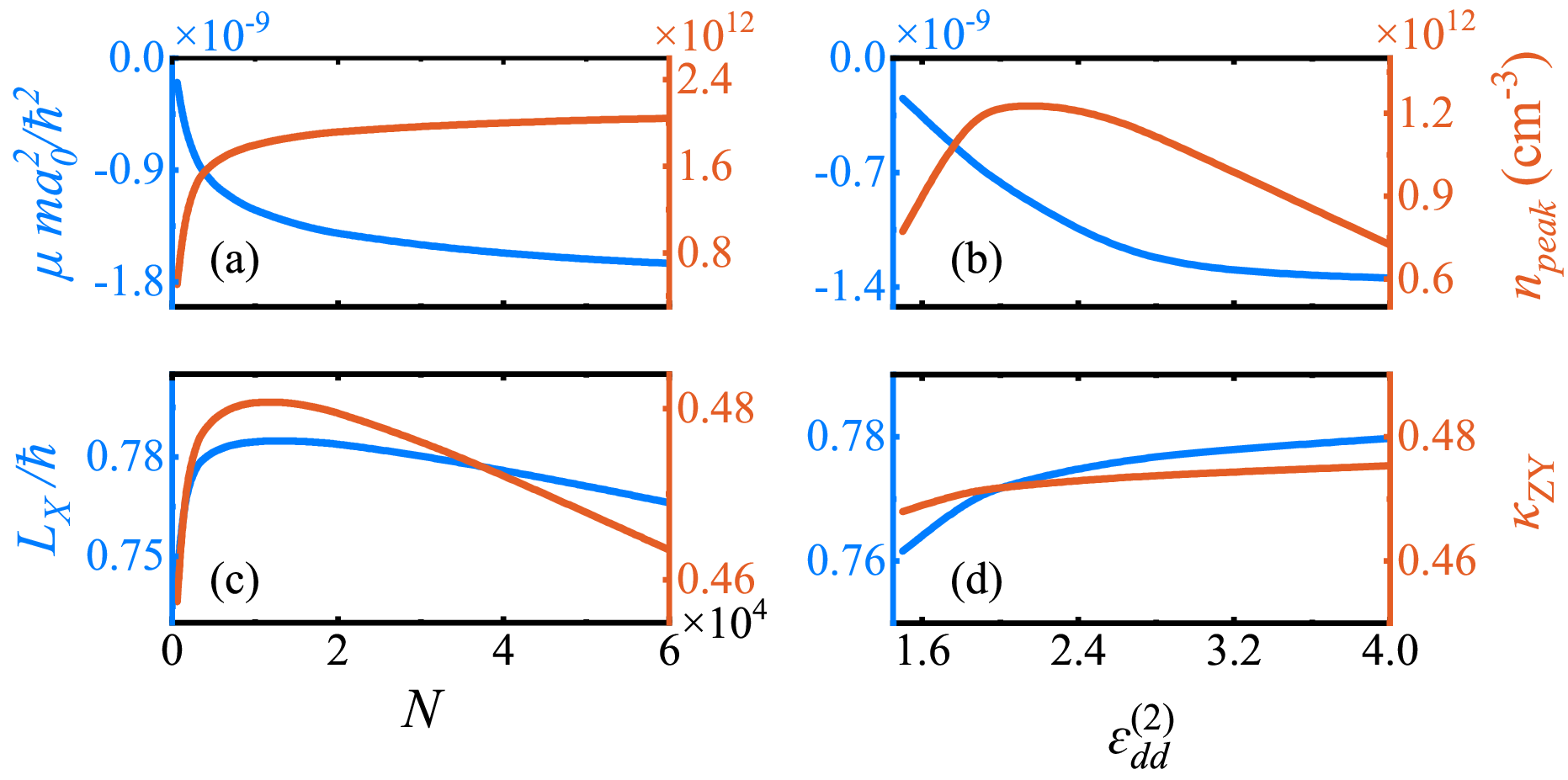}}
\caption{Chemical potential $\protect\mu$ and peak density
$n_{\text{peak}}$ as functions of the particle number $N$
in (a) and the DDI strength $\protect\epsilon_{\mathrm{dd}}^{(2)}$
in (b). The corresponding dependences of the normalized
angular momentum $L_X$ and spatial anisotropy $\protect\kappa_{\text{ZY}}$
are shown in (c) and (d), respectively. The $N$-dependent
results are obtained at fixed $\protect\epsilon_{\mathrm{dd}}^{(2)}=1.8$,
while the $\protect\epsilon_{\mathrm{dd}}^{(2)}$-dependent results
are obtained at fixed $N=1500$.}
\label{fig2}
\end{figure}

We consider solutions characterized by an integer winding number $S$, with
the vorticity axis oriented along the $X$ axis. Such states are numerically
generated by means of the imaginary-time propagation (ITP) method \cite{ITP}%
, initiated with the anisotropic ansatz of the form
\begin{equation}
\psi (X,Y,Z)=A\tilde{r}^{S}\exp (-\alpha _{1}X^{2}-\alpha _{2}Y^{2}-\alpha
_{3}Z^{2}+iS\tilde{\theta}),  \label{S-1-input}
\end{equation}%
whose amplitude and spatial widths are determined by positive constants $A$
and $\alpha _{1,2,3}$. The deformed polar coordinates in the $(Y,Z)$ plane
are defined as
\begin{equation}
(\tilde{r},\tilde{\theta})\equiv \left( \sqrt{Y^{2}+\beta ^{2}Z^{2}},\arctan
(\beta Z/Y)\right) ,  \label{beta}
\end{equation}
where $\beta >1$ is an anisotropy factor. It is found that, for fixed
constants $\left( N,\epsilon _{\mathrm{dd}}^{(2)}\right) $, varying $\beta $ in the
input (\ref{S-1-input}) does not change the final stationary solution,
affecting only the convergence rate of the ITP. The spatial structure of a
representative vortex solution with $S=1$, $N=1500$, and $\epsilon
_{dd}^{(2)}=1.8$ is presented in Figs.~\ref{fig1}(a)--\ref{fig1}(c). While
the three-dimensional (3D) profile exhibits a distinct anisotropic
pancake-like shape in (a), the corresponding cross sections in the $X=0$
plane clearly unveil the elliptical density profile [panel (b)] and the
associated phase structure in (c). This morphology is produced by the
anisotropic microwave-dressed DDI, which may have different signs in
different directions. As it follows from Eq.~(\ref{DDI}), or $\epsilon
_{\mathrm{dd}}^{(0)}=0$ and $\epsilon _{\mathrm{dd}}^{(2)}\neq 0$, the signs are opposite
along the $X$ and $Y$ directions, while $\epsilon _{\mathrm{dd}}^{(2)}$ vanishes
along the $Z$ axis, thereby producing the observed pancake-shaped density
distribution.

The robustness of these states against small perturbations was tested by
real-time simulations of the perturbed evolution, mapping the resultant
stability diagram in the $(N,\epsilon _{\mathrm{dd}}^{(2)})$ parameter plane (Fig.~%
\ref{fig1}(d)). The vortex stability region nearly coincides with that of
the ground state with $S=0$, demonstrating the robustness of the vortex
states and the suitability of the microwave-dressed molecular condensates
for exploring vortex physics.

To systematically characterize the structural properties of the $S=1$ vortex
QDs, we define their spatial anisotropy $\kappa
_{\text{ZY}}$ in the $X=0$ plane, the peak density $n_{\text{peak}}$, and
the normalized angular momentum $L_{X}$:%
\begin{equation}
\kappa _{\text{ZY}}\ =\frac{W_{Z}}{W_{Y}},\quad L_{\text{X}}=\frac{1}{N}\int \psi
^{\ast }\hat{L}_{\text{X}}\psi \,d\mathbf{R},  \label{kappa}
\end{equation}
where the effective widths along the $Y$ and $Z$ axes are
defined as $W_{\text{Y}}=\left( \int |\psi (0,Y,0)|^{2}dy\right) ^{2}/\int |\psi
(0,Y,0)|^{4}dy$ and $W_{\text{Z}}=\left( \int |\psi (0,0,Z)|^{2}dz\right) ^{2}/\int
|\psi (0,0,Z)|^{4}dz$, respectively, and $\hat{L}_{\text{X}}=-i\hbar (Y\partial
_{Z}-Z\partial _{Y})$ denotes the $X$-component angular momentum operator.

Figure~\ref{fig2}(a,c) presents these characteristics as functions of the
particle number $N$ for fixed $\epsilon _{\mathrm{dd}}^{(2)}=1.8$. As seen from the
blue curve in Fig.~\ref{fig2}(a), the chemical potential strictly obeys
the Vakhitov-Kolokolov criterion, $d\mu /dN<0$, which is the well-known
necessary stability condition \cite{Vakhitov1973,Berge} (although not
a sufficient one, for vortex states \cite{Malomed-vortex-review}). The full
stability of the vortex-QD family is corroborated by our real-time simulations of the
perturbed propagation. Note that, as $N$ increases, $\mu $ does not plateau
but rather keeps slow monotonous decline. The orange curve in Fig.~\ref{fig2}(a)
shows that the peak density increases continuously with the particle number,
without flattening out This
behavior is different from that expected for familiar QDs with the flat-top
structure \cite{Petrov,PA}, because the considered range of the particle
number, $N$, is insufficient to saturate the density along the tightly
confined $X$ axis. In fact, given the extreme spatial anisotropy of the
system, establishing a flat-top structure across all three dimensions would
demand an extraordinarily high value of $N$.

\begin{figure}[tbp]
{\includegraphics[width=1\columnwidth]{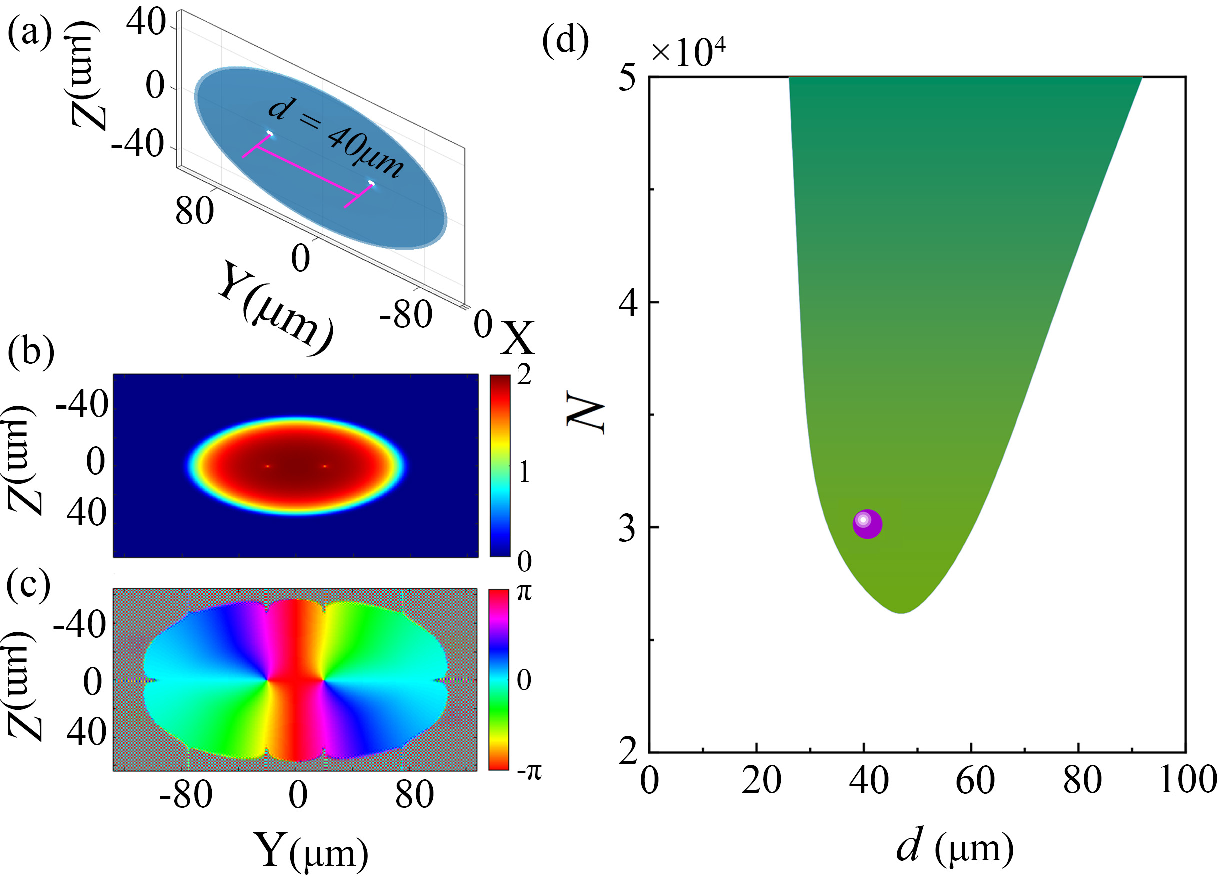}}
\caption{(a)--(c) A representative configuration of an anisotropic vortex QD
with $S=2$ \ and $(N,\protect\epsilon _{\mathrm{dd}}^{(2)},d)=(30000,1.8,40)$,
corresponding to the purple circle marked in the stability diagram (d). (a)
the 3D density isosurface; (b) and (c): the density and phase distributions
in the $X=0$ plane, revealing two well-separated vortex cores. (d) The
stability domain of the $S=2$ vortex QDs in the $(N,d)$ parameter plane,
where $d$ is the distance between the separated vortex pivots.}
\label{fig3}
\end{figure}

The anisotropy breaks the conservation of the angular momentum $L_{\text{X}}$ and
its direct proportionality to the topological charge $S$. On top of its
slight variation with $N$, the evolution of $L_{\text{X}}$ is tightly constrained
by the system's geometry. As shown in Fig.~\ref{fig2}(c), the
numerical data are well described, within the parameter range considered, by
the relation $L_{\text{X}}/\hbar\simeq1.6\,\kappa_{\mathrm{ZY}}$, with the prefactor $%
1.6$ different from its usual value $2$ for dipolar atomic BECs~\cite%
{GLi-3D}. The reduced prefactor is depends on the strong anisotropy of the
molecular BEC, which modifies the transverse vortex-flow field and its
contribution to the total angular momentum.

The orange curve in Fig.~\ref{fig2}(c) shows that the spatial anisotropy factor $\kappa _{%
\mathrm{ZY}}$ varies weakly with the increase of $N$, keeping values $\kappa
_{\mathrm{ZY}}\simeq 2.1$, which are close to but slightly different from
the initial anisotropy parameter $\beta =2$ in Eq. (\ref{beta}).

\begin{figure}[tbp]
{\includegraphics[width=1\columnwidth]{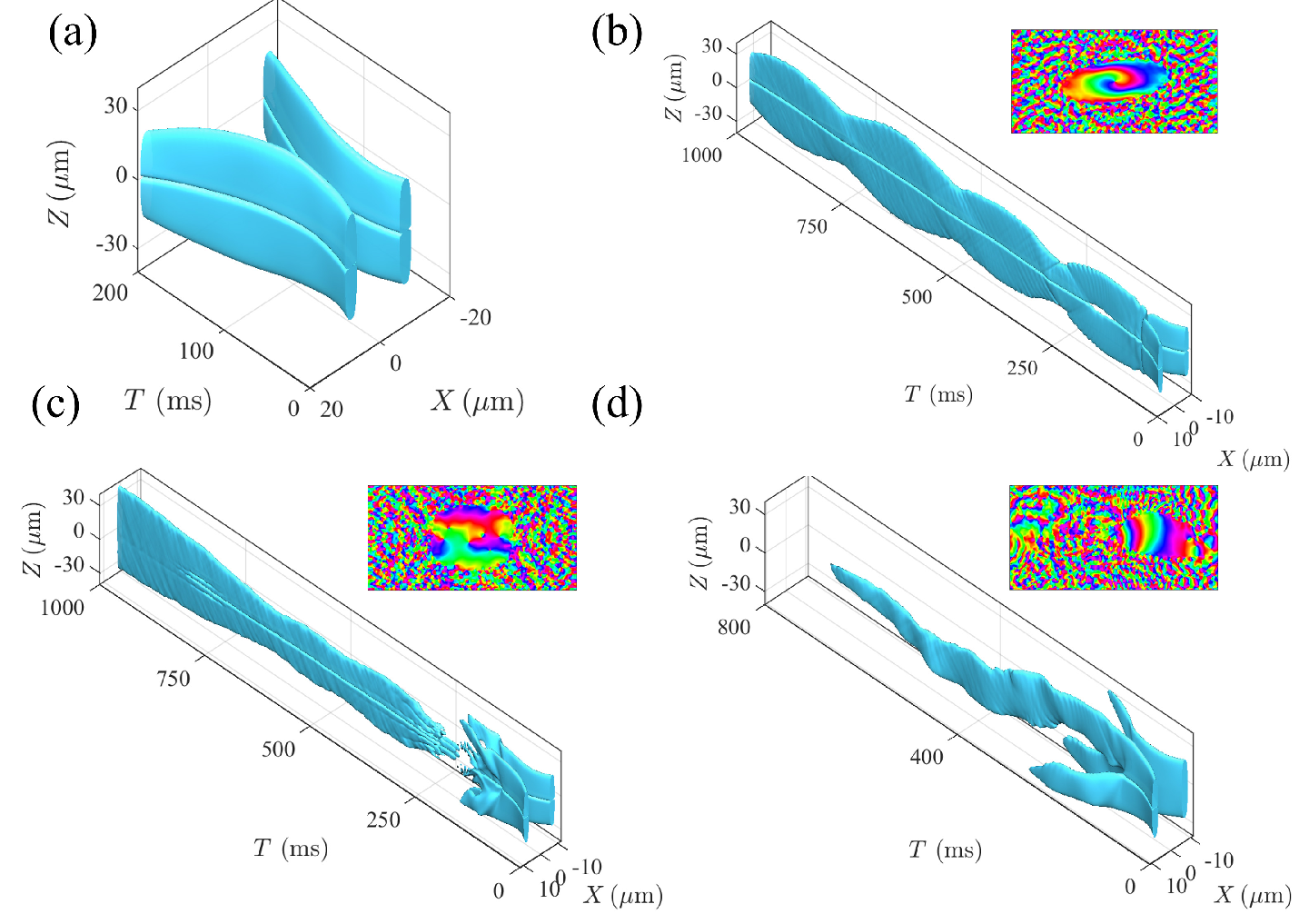}}
\caption{Collisions between QDs moving along the $X$ axis, initialized by $\psi (X,Y,Z,t=0)=\psi _{\mathrm{L}}(X+X_{0},Y,Z)\exp (i\eta X)+\psi _{\mathrm{R}}(X-X_{0},Y,Z)\exp (-i\eta X)$, where $\psi _{\mathrm{L}}$ and $\psi _{\mathrm{R}}$ are the stationary QDs centered at $X=\pm X_{0}$, and $\pm \eta $ are kicks setting them in motion, respectively. Collisions
between two identical vortex QDs with $S=1$, set in motion by kicks with $%
\protect\eta =\protect\pi /6$ (a) and $\protect\eta =\protect\pi /4$ (b). (c) The collision between two vortex QDs with
opposite topological charges, $S_{\mathrm{L}}=-S_{\mathrm{R}}=1$, and $%
\protect\eta =\protect\pi /4$. (d) The collision between the vortex QD with $%
S_{\mathrm{R}}=1$ and ground-state QD ($S=0$). at $\protect\eta =\protect\pi %
/4$. The upper-right insets in (b), (c) and (d) display the phase
distributions of the corresponding final states in $X=0$ plane. In all
cases, other parameters are $N=10000$ and $\protect\epsilon _{\mathrm{dd}}^{(2)}=1.8$.}
\label{fig5}
\end{figure}

Next, in Fig.~\ref{fig2}(b,d) we examine the same characteristics of the
vortex QDs as functions of $\epsilon _{\mathrm{dd}}^{(2)}$ at fixed $N=1500$. The
chemical potential decreases monotonously, while the the peak density first
increases and then decreases, attaining a maximum near $\epsilon
_{dd}^{(2)}=2$. The normalized angular momentum $L_{X}$ increases with $%
\epsilon _{\mathrm{dd}}^{(2)}$, while $\kappa _{\mathrm{ZY}}$ slightly decreases,
remaining close to its mean value $\approx 2.1$. Thus, $\kappa _{\mathrm{ZY}%
}\approx 2.1$ is a robust geometric characteristic of the stationary vortex
QDs.

To generate the anisotropic vortex state with $S=2$, we initialize the ITP
with an ansatz comprising two $S=1$ pivots placed at $Y=\pm d/2$ along the $Y
$ axis:
\begin{equation}
\begin{split}
\psi (X,Y,Z)=& A\left[ (Y-d/2)+i\beta Z\right] \left[ (Y+d/2)+i\beta Z\right]
\\
& \times \exp (-\alpha _{1}X^{2}-\alpha _{2}Y^{2}-\alpha _{3}Z^{2}),
\end{split}%
\end{equation}%
where constants $A$, $\beta $, and $\alpha _{1,2,3}$ play the same roles as
for $S=1$. In the established state, the separation between the two vortex
pivots remains close to $d$, which is imposed by the initial ansatz. Even if
the ITP is initialized with $d=0$, corresponding to the double vortex with
the single pivot, it converges to a stationary state with separated pivots.

Figures~\ref{fig3}(a)--\ref{fig3}(c) portray the spatial structure of a $S=2$
vortex state. As corroborated by the phase profile in the $X=0$ plane (Fig.~%
\ref{fig3}(c)), the system harbors two vortex pivots. Indeed, it is natural
that, under strong spatial anisotropy, the vortex with $S=2$ is prone to the
dissociation, splitting into two $S=1$ vortex cores with separation $d$
between them. The stability of such a two-core vortex QD strongly depends on
$d$, as seen in Fig.~\ref{fig3}(d). An increase in $N$ naturally increases
the spatial width of the solutions, thereby broadening the stability domain
for $d$. Note that the $S=2$ state is entirely unstable at $N<25000$ in Fig.~\ref%
{fig3}(d). The instability below a critical value of the norm is typical for
vortex solitons, with this value steeply growing with the increase of $S$
\cite{Malomed-vortex-review}.

%

Simulations of the evolution of QDs carrying vortex--antivortex pairs with
\(S=\pm1\) demonstrate that these pairs are unstable bound states. The VAV
pair fuses into a fundamental soliton, which then starts drifting along the
\(Z\) direction and eventually hits the boundary.
Generally, higher-order vortex QDs with $S>2$ tend to split into
singly-charged ones. Nevertheless, they are stable for sufficiently large
values of the particle number and separation between unitary vortex cores
inside of the droplet.

The dynamics of vortex droplets are investigated through two representative processes: collisions between moving vortex QDs and stability under a driven $\epsilon_{\mathrm{dd}}^{(0)}$.

Figure \ref{fig5} shows typical examples of head-on collisions between two vortex QDs moving in opposite directions along the $X$ axis. Two stationary QDs, initially centered at $X=\pm X_0$, are set into motion by kicks of $\pm\eta$, respectively. Slowly colliding QDs with $\eta<\eta_c$ (here $\eta_c\approx\pi/5$) undergo quasi-elastic rebound [Fig. \ref{fig5}(a)], whereas those with $\eta>\eta_c$ merge. In particular, two identical $S=1$ vortex QDs merge into a single $S=1$ state rather than an $S=2$ state [Fig. \ref{fig5}(b)], reflecting the nonconservation of angular momentum in the strongly anisotropic system. Collisions between oppositely charged vortex QDs ($S=\pm1$) result in fragmentation followed by fusion into a stable ground state [Fig. \ref{fig5}(c)]. By contrast, a collision between $S=1$ and $S=0$ QDs destroys the vorticity and produces a merged but dynamically unstable droplet, which eventually disintegrates [Fig. \ref{fig5}(d)].

\begin{figure}[tbp]
{\includegraphics[width=1\columnwidth]{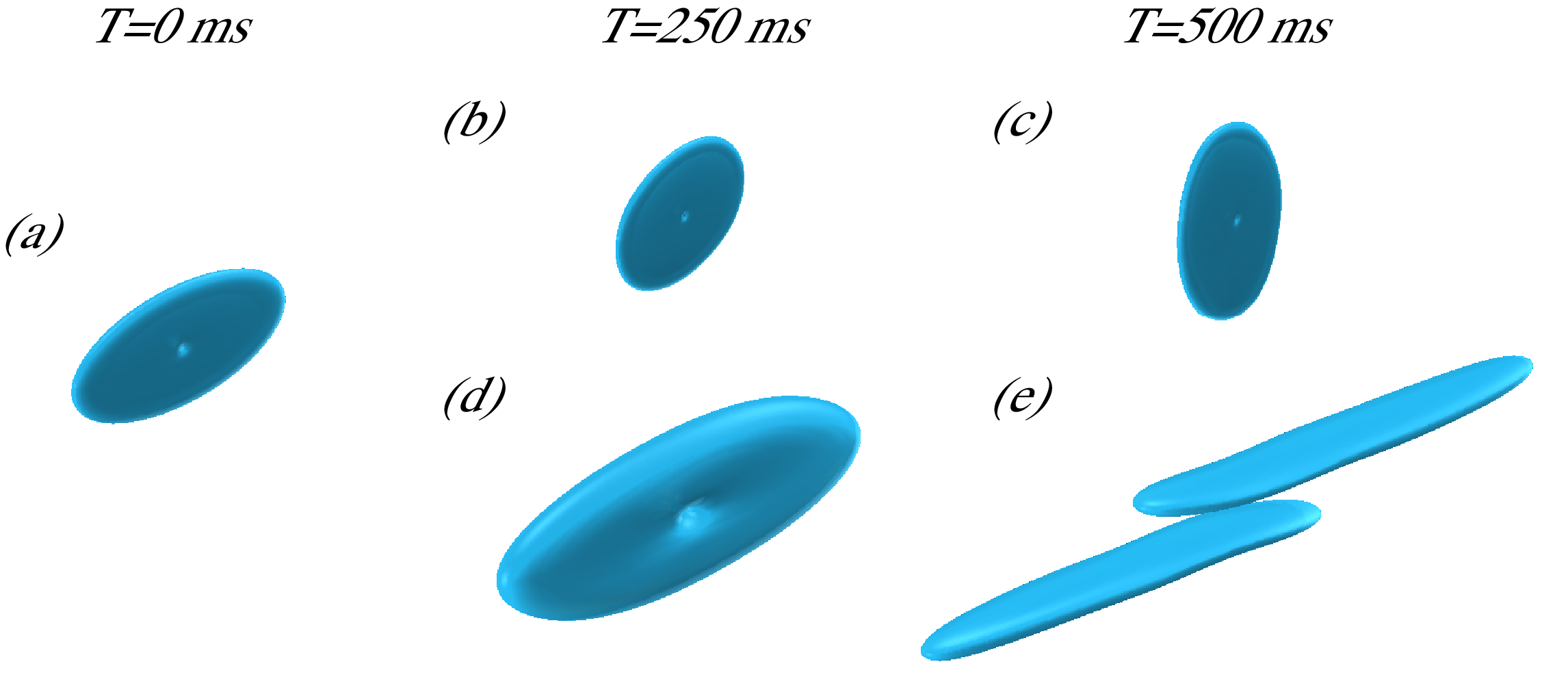}}
\caption{Stationary vortex QD with $(N, \epsilon^{(2)}_{\mathrm{dd}})=(2000,2)$ adiabatically driving by $\epsilon^{(0)}_{\mathrm{dd}}=\zeta T$ with $\zeta=\pm4\times10^{-3}$ ms$^{-1}$, respectively. Panel (a) shows the initial state at $T=0$. The upper sequence, panels (b) and (c), shows the evolution at
$T=250$ and $500\,\mathrm{ms}$ with $\zeta=4\times10^{-3}$ ms$^{-1}$, while
the lower sequence, panels (d) and (e), shows the evolution at the same times with $\zeta=-4\times10^{-3}$ ms$^{-1}$.}
\label{fig6}
\end{figure}

To examine the response of the vortex QD to $\epsilon_{\mathrm{dd}}^{(0)}$ [see Eq.~(\ref{DDI})], we adiabatically drive $\epsilon_{\mathrm{dd}}^{(0)}=\xi T$ from zero. Figure~\ref{fig6} shows the dynamics of a stationary vortex QD driven with $\zeta=\pm4\times10^{-3}$~ms$^{-1}$. During the drive, $\gamma_{\mathrm{LHY}}$ also changes accordingly through Eq.~(\ref{gamma-QF}). For $\zeta>0$, the vortex remains stable throughout the drive [Figs.~\ref{fig6}(a)--(c)]. In contrast, for $\zeta<0$, the droplet expands and elongates along $Y$, eventually splitting into two fragments as $\epsilon_{\mathrm{dd}}^{(0)}$ becomes sufficiently negative [Figs.~\ref{fig6}(a), (d), and (e)]. This contrast originates from the angular dependence of the DDI in Eq.~(\ref{DDI}). Along $Z$ ($\theta=0$), the angular factor is $-2\epsilon_{\mathrm{dd}}^{(0)}$. Positive $\epsilon_{\mathrm{dd}}^{(0)}$ therefore strengthens the attraction along $Z$, allowing the droplet to elongate in this direction while remaining self-bound. By contrast, negative $\epsilon_{\mathrm{dd}}^{(0)}$ makes the DDI repulsive along $Z$ and weakens self-binding, leading to expansion and fragmentation.

In conclusion, microwave-dressed molecular condensates provide a robust platform
for studying vortex quantum droplets (QDs) in strongly anisotropic quantum fluids.
The stability region of the singly charged ($S=1$) vortex nearly coincides with
that of the ground-state ($S=0$) QD, while higher-charge vortices can form stable
configurations with their multiply charged cores split into separated unit-charge
vortices. The stability of these split-pivot states depends on the system parameters
and pivot separation, and vortices with $S>2$ require sufficiently large particle
numbers and inter-core distances. Vortex-antivortex pairs are unstable and ultimately
fuse into a ground-state droplet. Angular momentum is not determined solely by
topological charge. For an $S=1$ vortex QD, its dependence on transverse anisotropy
is qualitatively similar to that in dipolar atomic BECs \cite{GLi-3D}, but with a
smaller coefficient, reflecting the stronger anisotropy of our system. In split-pivot
states, the total angular momentum depends on the separation of the inner pivots and
is not simply the sum of the contributions from the individual vortices. Collisions
between moving vortex QDs can lead to rebound, merger, or fragmentation, depending on
their topological charges and collision velocities. Driving $\epsilon_{\mathrm{dd}}^{(0)}$ away from zero reveals a sign-dependent response: the vortex remains stable for positive $\epsilon_{\mathrm{dd}}^{(0)}$, whereas negative $\epsilon_{\mathrm{dd}}^{(0)}$ induces expansion and eventual fragmentation. Overall, microwave control of anisotropic
dipole-dipole interactions offers a versatile means of engineering topological excitations
in molecular Bose-Einstein condensates.

These findings motivate extension of the analysis to for a molecular system
with external potentials, including strong ones which may reduce the
effective dimension from $3$ to $2$, thus strongly affecting the vortex
dynamics \cite{GLi-2D,Malomed-vortex-review,JDutta,HWang}. It is also
interesting to address the structure and dynamics of multi-vortex complexes,
including QD chains, cf. Refs. \cite%
{GLi-3D,CXL-sum-rule,LSantos-excitations,Blakie-axial,TPfau-scissors}.

\begin{acknowledgments}
We appreciate valuable discussions with Dr. D. Baillie (University of Otago, New Zealand) and Guilong Li (Nanjing University, China). This work was
supported by NNSFC (China) through Grants No. 12274077, No. 12475014,
Guangdong Basic and Applied Basic Research Foundation No. 2024A1515030131,
No. 2025A1515011128, No. 2023A1515110198, No. 2023A1515010770, the Research Fund of
Guangdong-Hong Kong-Macao Joint Laboratory for Intelligent Micro-Nano
Optoelectronic Technology through grant No. 2020B1212030010.
\end{acknowledgments}


\begin{thebibliography}{99}
\bibitem{SRc} According to Ref. \cite{Baillie}, $U_{dd}^{R_c}(\mathbf{k}%
)/s(kR_c)=\epsilon_{dd}^{(0)}(3\cos^2\theta_k-1)-\sqrt{3}\sin^2\theta_k\cos
2\varphi_k$, where $s(kR_c)=1+3(kR_c)^{-2}\cos(kR_c)-3(kR_c)^{-3}\sin(kR_c)$%
. Here $R_c$ is a spherical upper radial cutoff. The interaction is retained
for $r<R_c$ and omitted for $r>R_c$. Thus, in the cutoff-removal limit $%
R_c\rightarrow\infty$, at fixed $k$, $s(kR_c)\rightarrow1$, and $%
U_{dd}^{R_c}(\mathbf{k})\rightarrow U_{dd}(\mathbf{k})$.
\end{thebibliography}
\end{document}